nucl-ex/9703002
\documentstyle{elsart}
\begin{document}

\newcommand{\ddxhe}{$d\,d\to\, $\nuc{4}{He}$\,X$}
\newcommand{\dpoldxhe}{$\vec{d}\,d\to\, $\nuc{4}{He}$\,X$}
\newcommand{\ddetahe}{$d\,d\to\,$\nuc{4}{He}$\,\eta$}
\newcommand{\dpoldetahe}{$\vec{d}\,d\to\,$\nuc{4}{He}$\,\eta$}
\newcommand{\ddpihe}{$d\,d\to\,$\nuc{4}{He}\,$\pi^0$}
\newcommand{\etahef}{$\eta\,$\nuc{4}{He}}
\newcommand{\etahet}{$\eta\,$\nuc{3}{He}}
\newcommand{\hef}{\nuc{4}{He}}
\newcommand{\het}{\nuc{3}{He}}
\newcommand{\pdetahe}{$p\,d\to\,$\nuc{3}{He}$\,\eta$}
\newcommand{\pipetan}{$\pi^-\,p\to n\,\eta$}

\begin{frontmatter}
\title{$\eta$--Helium Quasi-Bound States}

\author[IPN]{N.~Willis},
\author[IPN]{Y.~Le~Bornec},
\author[CRN]{A.~Zghiche},
\author[LONDON]{C.~Wilkin},
\author[IPN,LNS]{R.~Wurzinger},
\author[CRN]{O.~Bing},
\author[LNS]{M.~Boivin},
\author[IPN]{P.~Courtat},
\author[LNS]{R.~Gacougnolle},
\author[CRN]{F.~Hibou},
\author[IPN]{J.M.~Martin},
\author[LNS]{F.~Plouin},
\author[IPN]{B.~Tatischeff},
\author[LNS]{J.~Yonnet}

\address[IPN]{Institut de Physique Nucl\'eaire, IN2P3--CNRS, 
F--91406 Orsay Cedex, France}
\address[CRN]{Centre de Recherches Nucl\'eaires, IN2P3--CNRS / Universit\'e
Louis Pasteur, B.P.28, F--67037 Strasbourg Cedex~2, France}
\address[LNS]{Laboratoire National SATURNE, F--91191 Gif-sur-Yvette Cedex, 
France}
\address[LONDON]{University College London, London WC1E 6BT, United Kingdom}

\begin{abstract}
The cross section and tensor analysing power $t_{20}$ of the \dpoldetahe{}
reaction have been measured at six c.m.~momenta, $10\leq \mbox{\rm p}_{\eta}
\leq 90$~MeV/c. The threshold value of $t_{20}$ is consistent with
$1/\sqrt{2}$, which follows from parity conservation and Bose symmetry. The
much slower momentum variation observed for the reaction amplitude, as
compared to that for the analogous \pdetahe{} case, suggests strongly the
existence of a quasi-bound state in the \etahef{} system and optical model fits
indicate that this probably also the case for \etahet.


\end{abstract}
\end{frontmatter}

The near-threshold data on the \pdetahe{} reaction \cite{ber88,may95} 
are remarkable for both their overall strength and energy variation. To 
quantify this, let us define a spin-averaged amplitude squared in terms of the 
unpolarised c.m.~cross section by
\begin{equation}
\label{1}
|f|^2 = \frac{p_{d}}{p_{\eta}}\,\left(\frac{d\sigma}{d\Omega}\right),
\end{equation}
where $p_{d}$ and $p_{\eta}$ are the initial and final c.m.~momenta
respectively. The threshold value for $\eta$ production is as big as that
for $\pi$ production at its threshold, despite the much larger momentum
transfer \cite{ber88,may95}. This has been interpreted as evidence for
two-step processes with intermediate virtual pions \cite{FW1}. Furthermore,
while remaining essentially isotropic, $|f|^2$ falls by over a factor of
three from threshold up to $p_{\eta}\approx 0.35$~fm$^{-1}$ \cite{may95}. As
this corresponds to the change of only a few MeV in the incident beam energy,
it must be associated with an interaction in the final state between the
$\eta$ and \het{} \cite{CW1}. Parametrising this interaction by an
$S$--wave scattering length formula,
\begin{equation}
\label{2}
f = \frac{f_B}{1-iap_{\eta}}\:,
\end{equation}
where $f_B$ is slowly varying, the best fit is obtained with a 
scattering length \cite{may95}
\begin{equation}
\label{3}
a = \pm (3.8\pm 0.6) + i(1.6\pm 1.1)\ \mbox{\rm fm.}
\end{equation}
It should be noted that the data are completely insensitive to the sign of
$Re\{a\}$ and moreover that the errors on the real and imaginary parts of 
the scattering length are strongly correlated in the fit.

Such a large scattering length raises the question \cite{CW1} of whether the
$\eta$ meson can form quasi-bound states with nuclei much lighter than those
originally suggested \cite{Liu}. The existence of such states depends
critically upon the {\it sign} of the real part of the scattering length, with
$Re\{a\} < 0$ corresponding to `binding'; ({\it c.f.} the spin-triplet and
singlet nucleon--nucleon systems where the negative value of $a_t$ corresponds
to the existence of the deuteron). This quantity is difficult to estimate {\it
ab initio}, due to both uncertainty in the form of the multiple scattering
schemes to be used \cite{CW2} and incomplete knowledge of the basic
$\eta$--nucleon input \cite{Svarc}. 

The analogous \etahef{} channel is most easily accessed through the 
\ddetahe{} reaction though, since the cross section is much smaller, 
the signal-to-background ratio is less favourable than for the 
\etahet{} case. The signal can be enhanced by using a tensor polarised 
deuteron beam since parity conservation for a system of two identical 
deuterons ensures that at threshold or in the forward direction only deuterons 
with helicity $m=\pm 1$ can initiate the reaction \cite{FW2}, leading to a 
tensor analysing power of $t_{20}=1/\sqrt{2}$. Constructing cross sections 
with different helicities through
\begin{equation}
\label{4}
\sigma_{0} = (1-t_{20}\sqrt{2})\, \sigma\:,\ \ \ \ \ \ 
\sigma_{\pm 1} = (1+t_{20}/\sqrt{2})\, \sigma\:,
\end{equation}
where $\sigma$ is the unpolarised cross section, any $\eta$ signal near 
threshold should be observed only in $\sigma_{\pm 1}$.

Unpolarised \ddetahe{} data have been published from the SPES~IV spectrometer
at the Laboratoire National SATURNE (LNS) \cite{fra94} for $p_{\eta}$ between
14 and 40~MeV/c. The present experiment used the large acceptance spectrometer
SPES~III at LNS. The \dpoldetahe{} reaction was measured over a wider range of
$p_{\eta}$, at six different beam energies $T_d$ above threshold between 1121
and 1139~MeV ($10<p_{\eta}<90$~MeV/c) with a tensor polarised deuteron beam.
Two additional measurements were performed below the $\eta$ threshold to study
background contributions. The SPES~III magnetic spectrometer is sensitive to
particles produced in a forward laboratory cone of $\pm 3^{\circ}$, giving a
large solid angle of $\Delta\Omega\approx 10$ msr. The momentum acceptance in
p/Z of $600-1400$~MeV/c covers the whole possible range of the \dpoldxhe{}
reaction and allows a full phase space acceptance $(4\pi)$ for $\eta$
production from threshold up to 1139~MeV. This is a clear advantage over the
SPES~IV experiment where, even close to threshold, only $5-18\%$ of events 
could be seen \cite{fra94}.

A cryogenic deuterium target of 169~mg/cm$^2$ thickness was used.
Empty-target measurements showed a flat background contribution of the order
of 10\%. The deuteron beam flux was monitored 
with an ionisation chamber in the beam and two scintillator telescopes viewing 
the target. Absolute normalisation of this monitoring system was established 
through the carbon activation method \cite{ban71}. After also taking into 
account uncertainties in the target thickness, angular acceptance, 
corrections for detection efficiency and dead-time, an overall accuracy in the 
cross section normalisation of $\pm 12\%$ was achieved. The beam tensor 
polarisation is known to be very stable with a value of 
$\rho_{20} = 0.649\pm 0.011$ \cite{Egle}.

Phase space varies as $\sqrt{T_d-T_d^{\mathrm{\, thresh}}}$ so that a good
knowledge of the beam energy is vital in evaluating the cross section very
close to the production threshold energy $T_d^{\mathrm{\, thresh}}$. The
nominal beam energy delivered by the SATURNE synchrotron is subject to slight
offsets of up to $1-2$~MeV, though changes of up to 20~MeV required in this
experiment can be controlled to better than $\pm0.1$~MeV. We observed both the
forward and backward c.m.~$\alpha$ peaks and close to the threshold of $\eta$
production their separation is a very sensitive determination of $p_{\eta}$. 
Taking the value of the $\eta$ mass as m$_{\eta}=547.45$~MeV/c$^2$ \cite{pdg96}
and introducing it into a two-body Monte Carlo program, the beam energy is the
only free parameter. Comparing data with the simulation fixes the deviation
from the nominal central beam energy to be $(-1.2\pm0.1)$~MeV. A slightly
different assumption for m$_{\eta}$ would give a modified energy shift, but
this would be of no consequence when looking at the cross section as a function
of $p_{\eta}$. A spread in beam energy arises mainly from the synchrotron
($\sigma_p/p \approx 2.5\times 10^{-4}$) with smaller contributions coming from
energy loss in the target, leading to an energy width of $\sigma \approx
250$~keV. 

The $\alpha$ particles were identified by time-of-flight and energy loss
measurements between two scintillator hodoscopes placed 3~m apart. The 
electronics were tuned such as to suppress online most of the other particles 
produced, so that about 20\% of the events written on tape contained
$\alpha$ particles. The position measurement and trajectory reconstruction
back to the target, and hence the evaluation of the momentum of the
$\alpha$ particles, were carried out using the information from two multiwire
drift chambers. Further experimental details on the SPES~III set-up can be 
found in ref.\cite{mpc91}.

Fig.~1 shows the $\alpha$ particle laboratory momentum spectrum for the
\ddxhe{} reaction at $T_d=1125.8$~MeV without using the polarisation
information of the beam. This is 5.5 MeV above the $\eta$ production threshold
of $T_d^{\mathrm{\, thresh}}=1120.3$~MeV. The most prominent feature of the
data are the so-called ABC enhancements \cite{abc63,ban76} of the two-pion
spectrum at $p/Z\approx 780$ and $1260$~MeV/c, corresponding to a missing mass
$m_{\mathrm{ABC}}\approx 310$~MeV/c$^2$ produced in the forward and backward
directions in the c.m. system. The maximum possible missing mass of the
unobserved particle $X$ is detected near the centre of the spectrum in fig.~1,
where a third broader peak is apparent.

The central peak contains events from two different reactions. There is first a
`bump', whose shape was fitted by a simple fourth order polynomial that is
shown in the figure as a broken line. This `bump' is also present below the
$\eta$ threshold and its shape does not change significantly throughout the
kinematic conditions of this experiment. Secondly, situated upon the central
`bump' is an almost rectangularly shaped peak which shows up for all
measurements above the $\eta$ threshold. The shape is typical of an $S$--wave
two-body reaction near threshold measured with complete spectrometer
acceptance. It can be unambiguously associated with the \ddetahe{} reaction.
The width of the peak is a direct measure of $T_d-T_d^{\mathrm{\, thresh}}$,
the rise on either side reflecting the resolution of beam and spectrometer.

To evaluate the number of \ddetahe{} counts one could simply apply the fitting
method shown in fig.~1. Though all the $\eta$ events are contained within the
central rectangle, it would be hard to evaluate the two-body cross section with
precision without a detailed understanding of the central `bump'. There are
however two important pieces experimental information which can improve
significantly the signal-to-background ratio and reduce the model dependence.

Transformation of the data in fig.~1 into the c.m. system makes use of the
experimental knowledge of the scattering angle
$\theta_{\alpha}^{\mathrm{cm}}$. The peak for the two-body $\alpha\eta$
production will therefore be narrow as compared to the slowly varying
$\alpha\pi\pi$ three-body reaction which dominates the background.
In an attempt to eliminate non-$\eta$ background, data taken just below 
threshold were subtracted from above-threshold results. In the central region
this could be done at fixed values of $p_{\alpha}$ but, by introducing 
a scaling variable
\begin{equation}
\label{5}
x = p_{\alpha}/p_{\alpha}^{\mathrm{max}}\:,
\end{equation}
it is possible to extend this comparison even into the ABC regions. Here
$p_{\alpha}^{\mathrm{max}}$ is the maximum possible value of the
$\alpha$ particle momentum, corresponding to the production a system with 
$m_x=2m_\pi$.

The other important information to be used is the beam polarisation. The data
transformed into the c.m.~system are shown for helicities $m=0$ and $m=\pm 1$
in fig.~2. The two upper plots show data for a beam energy below the $\eta$
threshold, $T_d=1116.7$ (figs.~2a$_0$ and 2a$_1$). The central `bump' is now
divided into two sharper structures with a central dip dictated by phase space.
Events where the $\alpha$ particle goes in the forward c.m.~hemisphere are plot
with positive $x$. It should be noted that in general the polarised cross
section for two-pion production need not to be symmetric around
$\theta^{\mathrm{cm}}=90^{\circ}$. Analogous data above threshold at
$T_d=1125.8$~MeV, shown in figs.~2b$_0$ and 2b$_1$, look very similar apart
from the presence of two narrow peaks in fig.~2b$_1$ corresponding to forward
and backward going $\eta$ mesons. Using the beam intensity monitors to
renormalise the below-threshold data to the same beam luminosity, it is
possible to subtract the spectra from the above-threshold data, and these are
shown in figs.~2c$_0$ and 2c$_1$. For helicity $m=0$ there is no sign of any
visible $\eta$ signal so that $t_{20}$ for $\eta$ production is consistent with
a value of $+1/\sqrt{2}$. This provides a quantitative test of our subtraction
procedure since conservation laws forbid such contributions. The two $\eta$
peaks for $m=\pm$1 sit on a random background which on average vanishes. The
number of good \etahef{} events is obtained by integrating the counts in the
peaks and this leads to statistical errors of the order of 2\%. The systematic 
error of the method has been carefully evaluated from the fluctuations  outside
the $\eta$ peaks and found to be between 3.5 and 8.3\% of the counting rate in
the peak, depending on the accumulated statistics.

The same procedure has been carried out on data from all the above-threshold
measurements. Since there is full phase space acceptance for $\eta$ events up
to $T_d=1139$~MeV, no assumptions have to be made regarding the angular
distribution of the reaction in order to extract the total production cross
sections, whose values are given in table~1. In addition to the systematic
errors discussed earlier, it must be noted that the lowest energy point
involves an extra uncertainty due to an imprecision
of 0.1~MeV in the central value of the beam energy and uncertainty in the shape
of the energy spread. There is therefore significant model dependence for this
point, especially in the value of the effective target thickness, induced 
through the energy loss in the target.

The spin-averaged $|f|^2$ are evaluated assuming the
angular distribution to be isotropic, as is the case for \pdetahe{} up to quite
high $\eta$ momenta \cite{may95}, and which is consistent with our limited
angular information. The values given in table~1 take into account that no
$S$--wave $\eta$ production is possible for $m=0$. The results of this and the
SPES~IV experiment \cite{fra94}, shown in fig.~3, are completely consistent
though it must be borne in mind that the lowest points of both our and the
SPES~IV data have large error bars since the two experiments are subject to
similar beam momentum resolution problems and hence effective target thickness
uncertainties.

The variation of $|f|^2$ with $p_{\eta}$ is markedly less steep than that 
found for the \etahet{} system \cite{may95} and whose data, also 
illustrated in fig.~3, have suggested the existence of a quasi-bound
\etahet{} state. 

Direct fitting of the data with the scattering length formula of eq. (2) leads
to large correlations between the real and imaginary parts and it is therefore
appealing to fit both the \etahet{} and \etahef{} data simultaneously
using a lowest order optical potential \cite{CW1}, for which
\begin{equation}
\label{6}
2m_R\,V_{\mathrm{opt}}(r) =-4\pi A\,\rho(r)\,a(\eta N)\:,
\end{equation}
where $m_R$ is the $\eta$--nucleon reduced mass and $A$ the mass number of the
residual nucleus. Gaussian nuclear densities $\rho(r)$ with rms radii of 
1.9~fm and 1.63~fm were assumed for the \etahet{} and \etahef{} systems 
respectively. The best overall agreement is found with an $\eta$--nucleon
scattering length of $a(\eta N) \approx (0.52+0.25i)$~fm. This is not
dissimilar to the value deduced from \pipetan{} \cite{Svarc}, but 
second order effects in the optical potential must be taken into account 
before a quantitative comparison is made.

This input $a(\eta N)$ scattering length leads to
\begin{eqnarray}
\nonumber
a(\eta\,^3\mbox{\rm He}) &\approx& (-2.3+3.2i)\:\mbox{\rm fm}\:,\\
\label{7}
a(\eta\,^4\mbox{\rm He}) &\approx& (-2.2+1.1i)\:\mbox{\rm fm}\:, 
\end{eqnarray}
and the resulting curves are shown in fig.~3 (full lines). The negative signs 
on the real parts indicate that {\em both} systems are `bound', though the
imaginary parts give widths overlapping the thresholds. The difference to the
fit given in eq.(3) reflects the error correlation between the real and
imaginary parts of $a$(\etahet) which makes it hard to fix both parameters
purely from the \etahet{} production data.

In fact, independent of any details of the $\eta$--nucleus scattering scheme, 
the \etahef{} system is expected to be more bound than \etahet{} due
to the smaller radius of the \nuc{4}{He} nucleus and the presence of one extra
nucleon. This implies that in any scattering length fit
\begin{equation}
\label{8}
\left|Re\left\{1/a\left(\eta\,^4\mbox{\rm He}\right)\right\}\right|
> \left|Re\left\{1/a\left(\eta\,^3\mbox{\rm He}\right)\right\}\right|\:,
\end{equation}
and this is valid for the parameters derived above ($0.36$ {\it versus}
$0.15$~fm$^{-1}$).

Also shown in fig.~3 is the prediction of the two-step model with an
intermediate virtual pion (broken curve) \cite{FW2}. It is in reasonable 
agreement with the overall normalisation of the data, though the momentum
dependence is too sharp, due to the form of the final state interaction
assumed.

We have clearly shown that the momentum dependence is significantly 
flatter in the \etahef{} as compared to the \etahet{} case and this must arise
because this system is more `bound' and the pole pushed further away from the
physical region. This argument depends weakly on any reaction mechanism but
does not itself prove that \etahet{} is `bound', though this is suggested
strongly by the optical potential analysis taken in conjunction with the
\etahef{} data. This indicates that further studies should be carried out on
the $\eta$--deuteron channel to seek an exotic dibaryon at the $\eta$ threshold
\cite{PFW}. On the other hand any $\eta$ interaction with a heavier nucleus is
likely to give too much binding to be seen in threshold production.

This was the last experiment carried out before the closure of the SPES~III
spectrometer. We should like to thank E. Tomasi-Gustafson for helping us in
the beam polarisation measurements. We also greatly acknowledge the SATURNE 
accelerator crew for the excellent beam quality and the technical support.

\newpage
\begin{figure}
\noindent
\caption{Unpolarised momentum spectrum from the \ddxhe{} reaction at
$T_d=1125.8$~MeV. Due to the large acceptance of SPES~III, events corresponding
to \ddetahe{} fall in the rectangular-shaped peak lying on top of the dashed
central `bump', whose shape is consistent with the data taken below the $\eta$
threshold. The large enhancements at $p/Z=800$ and 1250~MeV/c are known as the
ABC effect [14,15]. The missing mass scale [MeV/c$^2$] in the top part of the
figure is only valid for events with $\theta^{\mathrm{cm}}=0^{\circ}$ or
$\theta^{\mathrm{cm}}=180^{\circ}$. $\pi^0$ and $\eta$ masses are indicated.
While in the centre of the spectrum full phase space acceptance $(4\pi)$ is
provided, this no longer holds in the region of the ABC peaks.
\label{fig1}} 
\end{figure}

\begin{figure}
\caption{Spectra of the \dpoldxhe{} reaction with helicities $m=0$ (a$_0$,
b$_0$, c$_0$) and $m=\pm 1$ (a$_1$, b$_1$, c$_1$), in terms of the c.m.~scaling
variable $x=p_{\alpha}/p_{\alpha}^{\mathrm{max}}$. The correctly normalised 
data at $T_d=1116.7$~MeV~(a) are subtracted from those at $T_d=1125.8$~MeV~(b) 
to reveal two narrow peaks in (c$_1$) corresponding to forward and backward
$\eta$ production in the c.m.~system for helicities $m=\pm 1$. The $m=0$
subtracted spectrum shown in (c$_0$) is consistent with random background with
zero mean in the $\eta$ region.
\label{fig2}}
\end{figure}

\begin{figure}
\caption{Averaged squared amplitudes of the \ddetahe{} reaction as functions
of the $\eta$ c.m.~momentum. Closed circles represent the results of this
work; open circles are from ref.~[9]. Error bars shown represent the systematic
errors, including those arising from eq.~(1) due to the imprecision in 
$p_{\eta}$. The data of ref.~[2] on the \pdetahe{} reaction (crosses) vary
markedly faster with $p_{\eta}$. Both shapes are well reproduced with the
scattering length formula of eq.~(2) (solid curves) using the optical
potential fit values of eq.~(7), as explained in the text. The broken curves 
are predictions in the two-step model of ref.~[8], including a normalisation
factor of 2.5 in the \etahet{} case. 
\label{fig3}}
\end{figure}

\begin{table}
\caption{Total cross section for the \dpoldetahe{} with helicities $m=\pm 1$ as
functions of the c.m.~momentum $p_{\eta}$ and the mean deuteron laboratory beam
energy $T_d$, the scale for the latter being set using 
m$_{\eta}=547.45$~MeV/c$^2$~[14]. Since our data are consistent with the result
that only deuterons with $m=\pm 1$ can produce $S$--wave $\eta$ mesons, the
unpolarised cross section $\sigma= \mbox{${\textstyle
\frac{2}{3}}$}\,\sigma_{\pm1}$. The spin-averaged amplitudes squared $|f|^2$
are obtained by assuming the angular distributions to be isotropic. The
statistical errors of about 2\%  are added quadratically to the systematic
error of the subtraction method.  The lowest momentum point is subject to an
additional systematic error due to the imprecision in the
beam energy of 0.1~MeV causing an uncertainty in the effective target
thickness. Note that, following eq.(1), the  error in $|f|^2$ includes a
contribution from $\Delta p_{\eta}$. Not shown in the table is the overall
uncertainty of $\pm 12\%$ in absolute normalisation. }\vspace{3mm} 
\phantom{x}

\begin{tabular}{ccccccc}
\hline
T$_d$ &
$p_{\eta}$ &
$\Delta p_{\eta}$ &
$\sigma_{\pm1}$ &
$\Delta \sigma_{\pm1}$ &
$|f|^2$ &
$\Delta |f|^2$\\
(MeV) &
(fm$^{-1}$) &
(fm$^{-1}$) &
(nb) &
(nb) &
(nb/sr) &
(nb/sr)\\
\hline
1120.7 & 0.05 & 0.005 &  7.1 & $^{+1.5}_{-0.9}$ & 39.7 & $^{+15.3}_{-7.8}$\\
1121.8 & 0.12 & 0.008 & 13.2 & 0.9 & 31.2 & 3.0 \\
1122.6 & 0.15 & 0.006 & 14.6 & 0.6 & 27.0 & 1.5 \\
1125.8 & 0.24 & 0.004 & 19.2 & 1.6 & 22.3 & 1.9 \\
1131.8 & 0.37 & 0.002 & 20.6 & 1.5 & 15.6 & 1.1 \\
1138.8 & 0.46 & 0.002 & 22.4 & 1.6 & 13.6 & 1.0 \\
\hline
\end{tabular}
\end{table}


\begin{thebibliography}{99}
\bibitem{ber88} 
J.~Berger et al., Phys. Rev. Lett. 61 (1988) 919.
\bibitem{may95}
B.~Mayer et al., Phys. Rev. C 53 (1996) 2068. 
\bibitem{FW1} 
G.~F\"aldt and C.~Wilkin, Nucl. Phys. A 587 (1995) 769.
\bibitem{CW1} 
C.~Wilkin, Phys. Rev. C 47 (1993) R938.
\bibitem{Liu} 
R.S.~Bhalerao and L.C.~Liu, Phys. Rev. Lett. 54 (1985) 865.
\bibitem{CW2} 
C.~Wilkin, Physics Letters B 331 (1994) 276.
\bibitem{Svarc} 
M.~Batini\'{c}, I.~\v{S}laus and A.~\v{S}varc, 
Phys. Rev. C 52 (1995) 2188.
\bibitem{FW2} 
G.~F\"aldt and C.~Wilkin, Nucl. Phys. A 596 (1996) 488.
\bibitem{fra94} 
R.~Frascaria et al., Phys. Rev. C 50 (1994) R537;\\
F.~Roudot, Ph.D. thesis, IPN Orsay 1995, unpublished. 
\bibitem{ban71}
J.~Banaigs et al., Nucl. Inst. Meth. 95 (1971) 1479.
\bibitem{Egle} J.~Arvieux et al., Nucl. Inst. Meth. A 273 (1988) 48; \\
E.~Tomasi-Gustafson et al., Internal Report LNS/PH/91-27 (unpublished);\\
J.~Arvieux et al., Internal Report LNS/PH/92-24 (unpublished).
\bibitem{pdg96}
Particle Data Group, R.M.~Barnett et al., Phys. Rev. D 54 (1996) 1.
\bibitem{mpc91} 
M.P.~Combes-Comets et al., Phys. Rev. C 43 (1991) 973; \\
E.~Aslanides et al., Nucl. Phys. A528 (1991) 608.
\bibitem{abc63}
A. Abashian, N.E. Booth, K.H. Crowe et al., Phys. Rev. 132 (1963) 2296.
\bibitem{ban76}
J.~Banaigs et al., Nucl. Phys. B 105 (1976) 52.
\bibitem{PFW} F.~Plouin, P.~Fleury, and C.~Wilkin, Phys. Rev. Lett.
65 (1990) 690.
\end{thebibliography}
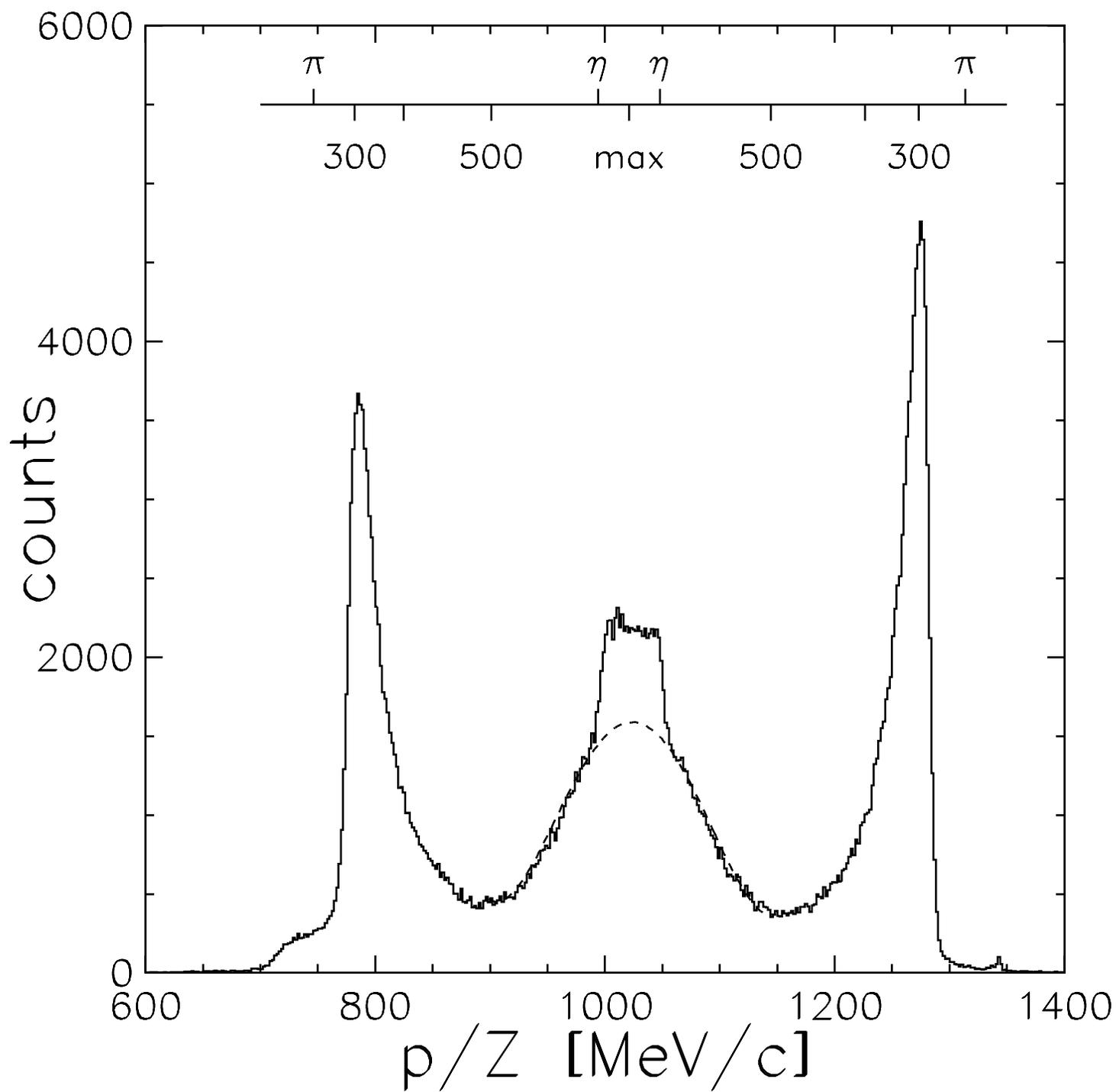
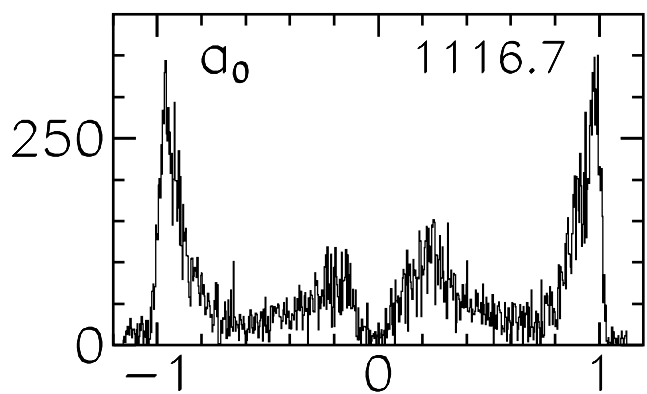
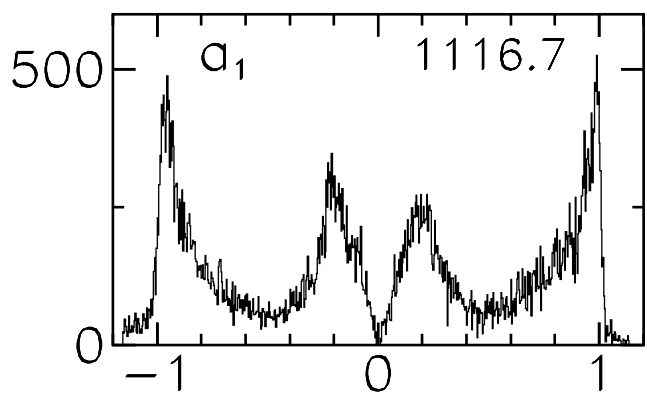
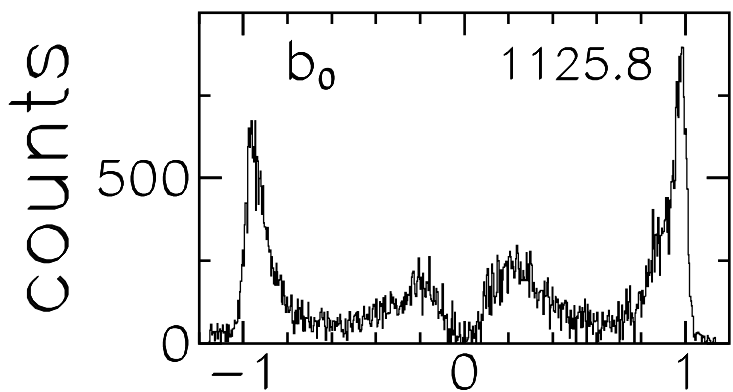
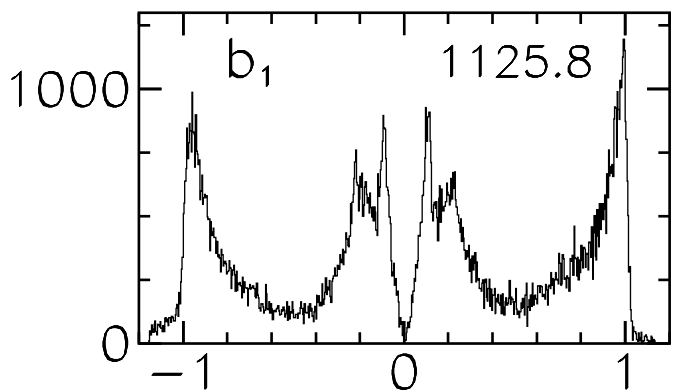
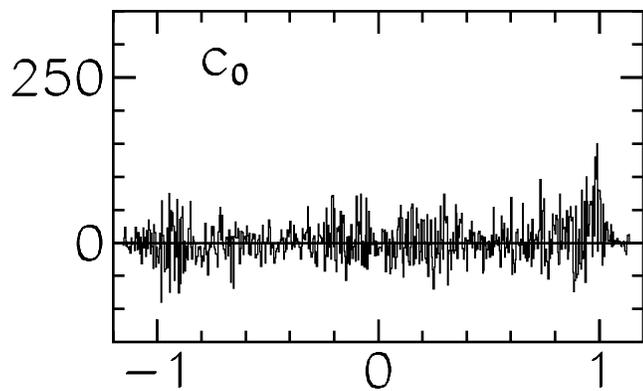
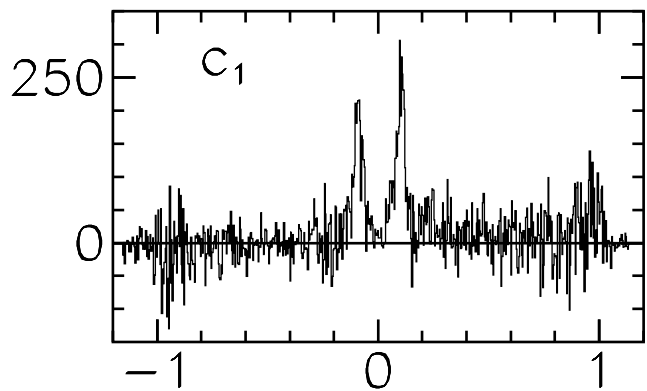
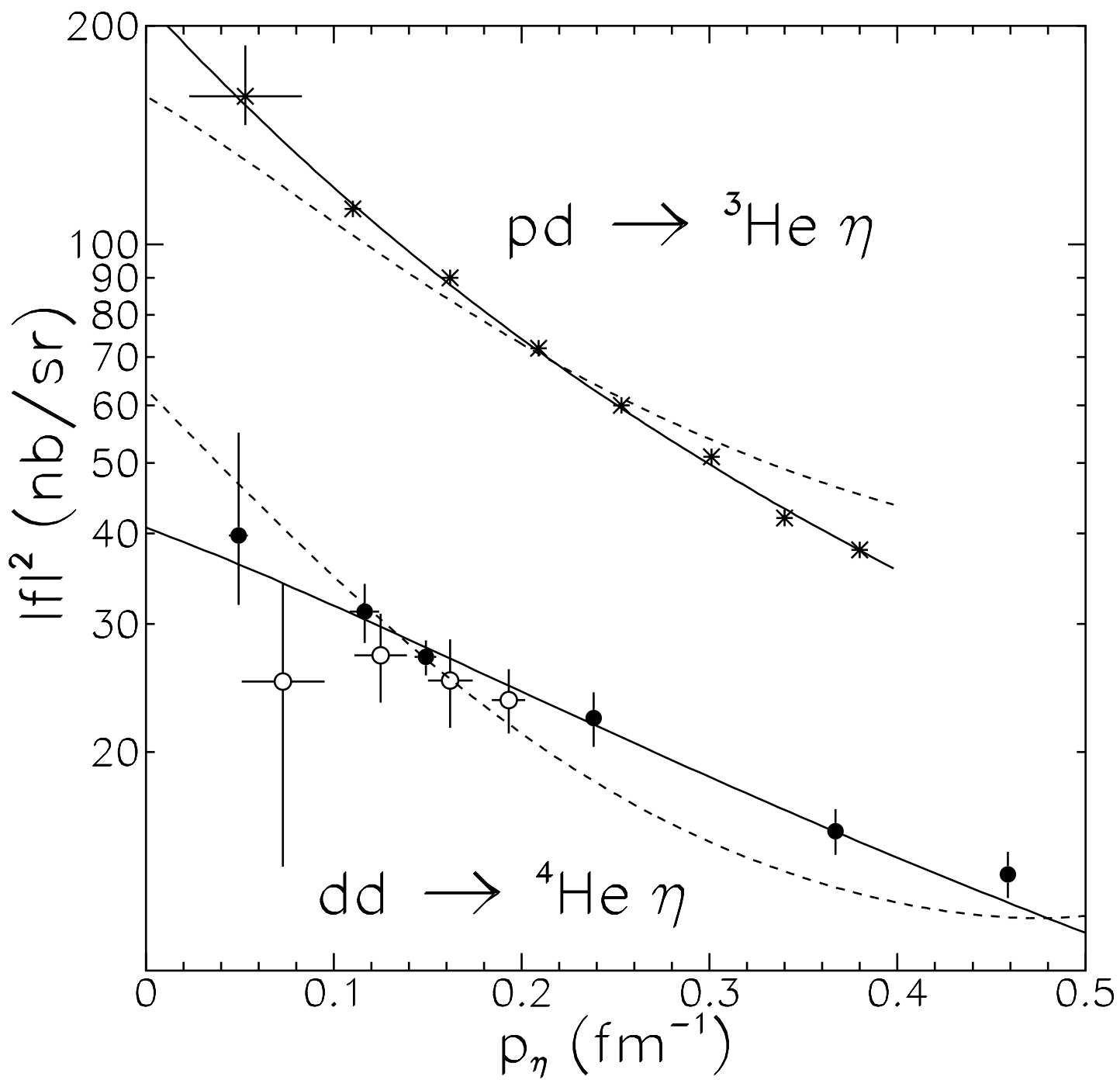
\end{document}